\documentclass[journal,
               10pt,
               letterpaper]{IEEEtran}

\ifCLASSINFOpdf
  \usepackage[pdftex]{graphicx}
  \graphicspath{{Figures/pdf/}{Figures/jpeg/}}
  \DeclareGraphicsExtensions{.pdf,.jpeg,.png}
\else
  \usepackage[dvips]{graphicx}
  \graphicspath{{../eps/}}
  \DeclareGraphicsExtensions{.eps}
\fi
\usepackage[cmex10]{amsmath}

\begin{document}
\title{Soft Decision Decoding of the Orthogonal Complex MIMO Codes for Three and Four Transmit Antennas}

\author{Risto~J.~Nordman,~\IEEEmembership{Member,~IEEE,}

\thanks{Manuscript received December 9, 2009. This work was supported in part by the European Commission under the project EUWB.}
\thanks{R. J. Nordman is with the VTT Technical Research Centre of Finland, Kaitov\"{a}yl\"{a} 1, FI-90571 Oulu, Finland (corresponding author to provide phone: \mbox{+358-40-704-5734;} fax: \mbox{358-20-722-2320;} e-mail: \mbox{risto.nordman@vtt.fi).}}}

\markboth{IEEE Transactions on Information Theory,~IT Paper 9-1014, Submitted 9~Decemeber~2009 as a regular paper}%
{Shell \MakeLowercase{\textit{et al.}}: Bare Demo of IEEEtran.cls for Journals}
%


\maketitle

\begin{abstract}
Orthogonality is a much desired property for MIMO coding. It enables symbol-wise decoding, where the errors in other symbol estimates do not affect the result, thus providing an optimality that is worth pursuing. Another beneficial property is a low complexity soft decision decoder, which for orthogonal complex MIMO codes is known for two transmit (Tx) antennas i.e. for the Alamouti code. We propose novel soft decision decoders for the orthogonal complex MIMO codes on three and four Tx antennas and extend the old result of maximal ratio combining (MRC) to cover all orthogonal codes up to four Tx antennas.

As a rule, a sophisticated transmission scheme encompasses forward error correction (FEC) coding, and its performance is measured at the FEC decoder instead of at the MIMO decoder. We introduce the receiver structure that delivers the MIMO decoder's soft decisions to the demodulator, which in turn cranks out the logarithm of likelihood ratio (LLR) of each bit and delivers them to the FEC decoder. This makes a significant improvement on the receiver, where a maximum likelihood (ML) MIMO decoder makes hard decisions at a too early stage. Further, the additional gain is achieved with stunningly low complexity.
\end{abstract}


%
\IEEEpeerreviewmaketitle

%
\section{Introduction}
%
\IEEEPARstart{M}{ultipath} propagation is a key feature of wireless channels. It has traditionally been conceived of as a bottleneck that hampers the creation of efficient wireless communication systems. It is a longstanding problem that has been addressed with several relatively effective solutions e.g. frequency hopping in GSM and spreading of the radio spectrum in UMTS. One well established method involves using multiple antennas on the receiver (Rx) side. In 1959 D.G. Brennan proposed an algorithm to combine the received constituent signals so that the signal-to-noise power ratio (SNR) of the compound signal becomes the sum of the constituent SNRs at the multiple Rx antennas. In the seminal paper~\cite{Brennan}, Brennan also proved the intuitive assumption that this is the highest achievable compound SNR in the case of uncorrelated additive noise. Brennan called the algorithm `maximal ratio combining' (MRC), and since it is optimum in single-in-multiple-out (SIMO) transmission systems, it can be considered a benchmark in comparison to more recently proposed multiple-in-multiple-out (MIMO) systems. We call the highest attainable compound SNR the `MRC limit', which is evidently an important optimality criterion in an MIMO receiver.

The era of MIMO coding really took off after the publication of the Alamouti code in 1998~\cite{Alamouti}. It is a rate 1 orthogonal code for two Tx antennas and --– as shown by
Alamouti --– provides the same diversity gain per antenna link as MRC i.e. $1\times2$ SIMO with MRC has the same diversity order as $2\times1$ multiple-in-single-out (MISO) with the Alamouti code. The Alamouti code is a perfect code, soft decoding of which is simple. Moreover, it is straightforward to show that in the case of the Alamouti code, the combiner of the signals from multiple Rx antennas reaches the MRC limit. Since it has such beneficial properties it is easy to understand the popularity of the Alamouti code. In fact, it is not so easy to surpass the Alamouti code's performance and still preserve low complexity in an MIMO scheme. Thus, it can be seen as another benchmark.

The Alamouti code was introduced as a space-time (ST) code, but we prefer using the more general term `MIMO code'. This is because the codes themselves do not prevent coding over frequencies and the term MIMO code applies to both ST and space-frequency (SF) coding as well as to their combination space-time-frequency (STF) coding. More generally, it is a question of diversity in terms of the number of branches from the Tx to the Rx side of MIMO; preferably these branches have a low correlation with one another. Whether the diversity reflects space, time or frequency is irrelevant for an MIMO code.

The invention of the Alamouti code triggered a search for other orthogonal MIMO codes, preferably applying to more than two Tx antennas. An extensive study of them was published as early as in 1999 by Tarokh et al.~\cite{TarokhTIT}, who proved that for complex constellations, the Alamouti code is the only rate one orthogonal code; for three and four Tx antennas, there are rate 3/4 codes. Further, Tarokh et al. proved the existence of complex MIMO codes with a rate of 1/2 for any number of Tx antennas. Their simulation results presented in~\cite{TarokhJSac} reveal that the orthogonal rate 3/4 codes can outperform the Alamouti code despite their lower channel data rate. Hence, they clearly have a larger potential for diversity gain. A closer look at~\cite{TarokhJSac}, however, gives rise to some questions as well. Specifically, Tarokh et al. apply the maximum likelihood (ML) decoder, which --– despite being optimum in that it cranks out the most likely set of transmitted constellation points --– is unable to pass any reliability
information to the FEC code decoder. This is not a problem in a theoretical study that omits the FEC code, but in the real world, the hard decisions at the FEC decoder input often reduce the achievable coding gain by 2 to 2.5 dB. This is a significant loss of ML decoding if it takes place at a very early stage of the receiver, since an FEC code is an essential part of any wireless digital transmission system. There was no FEC code in~\cite{TarokhJSac}, and the theoretical results of this eminent paper should be interpreted carefully when a real transmission system is a matter of concern.

An exhaustive study of the orthogonal codes was published by Liang in 2003~\cite{Liang}, where some methods to create high rate orthogonal complex MIMO codes were given. Furthermore, the generator matrices for the rate 3/4 codes for three and four Tx antennas were written on the form, which include only complex symbols and their conjugates. The quest for the maximum rate of the orthogonal complex MIMO codes for any number of Tx antennas was finally resolved in ~\cite{Liang}, where it was proven that for any even $2m$ and odd $2m-1$ number of Tx antennas the largest possible rate is $\left(m+1\right)/2m$. An equivalent bound was also presented by Wang et Xia~\cite{Wang} in a paper that was published in parallel with~\cite{Liang}. The bound in Wang et Xia was written lucidly as
\begin{equation}
	r \leq \frac{\left\lceil N_\text{Tx}/2\right\rceil+1}{2\left\lceil N_\text{Tx}/2\right\rceil}
	\label{eq:WangBound}
\end{equation}
where
$N_\text{Tx}$
is the number of transmit antennas and
$\left\lceil x \right\rceil$
stands for the ceiling function, i.e. the smallest integer $\geq x$.
Wang et Xia also introduced the concept of generalised orthogonal codes, where the rate was upper bounded by 4/5. This bound is loose in order to see if competitive generalised orthogonal complex MIMO code designs will be found in the future.

Orthogonality is a desired property, but it does not necessarily yield the best performance of the MIMO codes –-- especially when the measure is the error rate at the MIMO decoder output. Even the Alamouti code can be beaten by a sophisticated algebraic code, which modifies the constellation in a way that maximises the distances between the transmitted symbol sets~\cite{DamenNumT},~\cite{Belfiore}. Furthermore, there are full rate algebraic MIMO codes for multiple Tx antennas that guarantee great distances between the symbol sets, thus attaining amazingly low error rates at the MIMO decoder output,~\cite{Oggier}. Nonetheless, they suffer from two types of drawbacks, which are both related to the decoding. The ML decoder of an MIMO code finds the most likely set of transmitted symbols, but in the same context it is hard to extract any reliability information, let alone bitwise soft decisions. This is tantamount to saying that an MIMO code with a slightly more modest ML decoding performance can be highly competitive in cases where soft decision MIMO decoding and FEC code's performance are considered.
Besides, ML decoding of an algebraic MIMO code is a toilsome operation – despite the obvious relief provided by the sphere decoding algorithm~\cite{Viterbo}. The decoding complexity is a
real problem if the algebraic MIMO codes were meant to be used for very high data rate transmission.

The soft decision MIMO decoding is covered to some extent in existing literature. By combining the MIMO detector and the FEC decoder, Larsson \emph{et al}.~\cite{Larsson} strives for overall optimality and covers quite a lot of the theoretical aspects of the problem. Their mathematical model is relatively complex, however, which in turn explains why their numerical examples are limited to BPSK and $\left(7,3\right)$ Hamming code. The tractability of more powerful MIMO codes in combination with the variety of constellation mappings and FEC codes is still an issue, which~\cite{Larsson} --– despite its theoretical significance --– does not solve.

In the current paper, we study the bitwise soft information and show how it can be extracted and delivered to the FEC decoder --– with a startlingly low complexity though still in a nearly optimum way. As the measure of reliability, we first take bitwise LLR, from which we develop the vector LLR metric that contains the sufficient statistic for FEC decoder so as to be able to find the ML code word. In the process, we need a demodulator that directly calculates the bitwise LLR values instead of just the nearest constellation points and possibly their probabilities. It turns out that suboptimum solutions can be very simple, and we shed light on the basic means to attain the low complexity without forfeiting performance.

The key point of the proposed approach is the MIMO decoder and the combining of the signals received by multiple antennas. It is very important for the MIMO combiner not to lose information and for this reason we reproduced some analysis of the MRC limit achieving combiners for the $1\times N_\text{Rx}$ and $2\times N_\text{Rx}$ orthogonal MIMO codes i.e. for MRC and the Alamouti code. Further, we show how those combining algorithms can be extended to the orthogonal
MIMO codes for three and four Tx antennas as well. Hence, we introduce the soft decision signal combiners for the rate 3/4 orthogonal complex MIMO codes to reach the MRC limit, also preserving the low complexity structure of the rate 1 MIMO combiners. The novel combiners give us the tools to achieve and additional gain of 2 to 3 dB, which in turn renders the orthogonal MIMO codes for three and four Tx antennas much more competitive in systems with an FEC code.

The outline of the paper is as follows. In Section II, we discuss our MIMO transmission model, and its mathematical presentation. Section III contemplates the best form for the soft decisions to be passed from one module to another. We shed light on LLR and discuss its optimality. The combining algorithms of MRC and the Alamouti code are presented as examples to demonstrate the behaviour of the noise variance in the process. In Section IV, we present the novel soft decision decoding and combining algorithms for the orthogonal complex MIMO codes for three and four Tx antennas, and simulation results of their performances are shown in Section V. Finally, in Section VI we summarise the results.

%
%
\section{The MIMO Transmission Model}
For the mathematical tractability, we require that the channel response between the $\text{Tx} - \text{Rx}$ antenna pairs can be expressed as a complex scalar value. In the case of the impulse response, this means flat fading channels, which aren't normally encountered in practice. Therefore, we might prefer considering the frequency responses instead. The underlying system is thus essentially OFDM, though we do not specify any OFDM symbol or frame structure in this context. Instead, we apply the idealised Rayleigh fading channel model, which does not take into account the variety of sub-tones and other OFDM specific issues.

\begin{figure}
	\centering
	\includegraphics[width=3.4in]{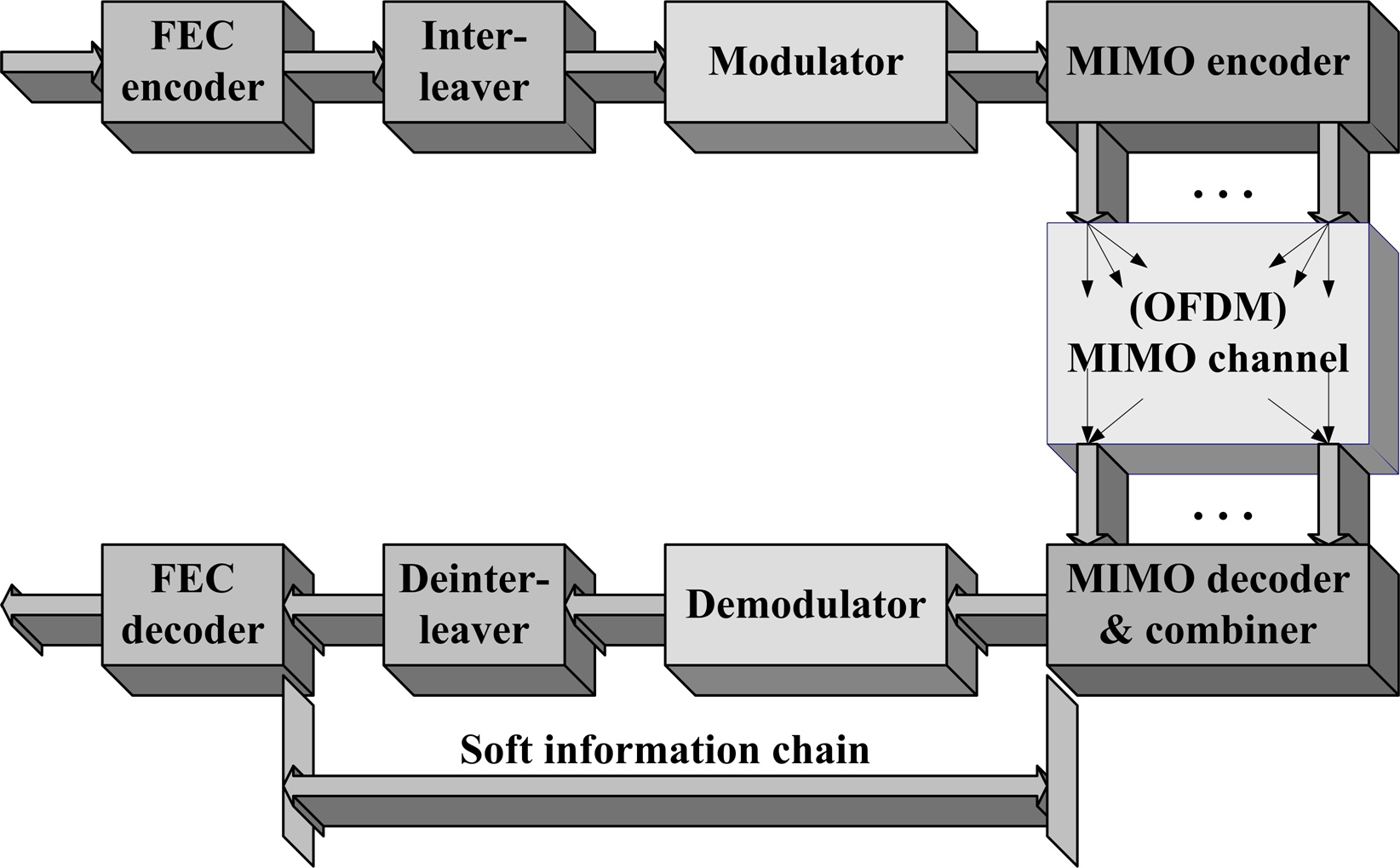}
 	\caption{The system model and soft decision flow in the receiver.}
	\label{fig:SystemModel}
\end{figure}

The model is depicted in Fig.~\ref{fig:SystemModel}, where the (OFDM) MIMO channel block in the middle represents the MIMO channel combined with the underlying OFDM processing. The soft information that is obtained in the MIMO decoder and combiner is transformed into bitwise LLR in the demodulator and then delivered to the FEC decoder via deinterleaver. The part of the receiver where the signal flow consists of soft decisions is indicated by the soft information chain arrow, and it forms the essence of our MIMO receiver.

\subsection{Mathematical Presentation}
We refer to the MIMO symbol as the set of constellation points that are transmitted simultaneously from $N_\text{Tx}$ antennas in a single instant. The MIMO code word consists of $l$ consecutive MIMO symbols. Let us consider the transmission
of one MIMO code word, which is composed of the information bearing constellation points $S_k$. It is transmitted in $l$ consecutive instants from $N_\text{Tx}$ and received by $N_\text{Rx}$ antennas. Whether the instants are adjacent in time or frequency is unimportant in the mathematical modelling. What is important is the set of channel responses, which must remain constant or at nearly constant to justify the approximation $H_{ij}\left(m\right)\approx H_{ij}\left(m+1\right)\approx H_{ij}\left(m+l+1\right)$,
where $i$ and $j$ stand for the Tx and Rx antenna indices respectively. The index $m$ refers to the starting point of the MIMO code word in the transmission chain. The samples received by $N_\text{Rx}$ antennas during $l$ consecutive instants are expressed as
\begin{equation}
	\mathbf{R}=\mathbf{GH}+\mathbf{W}
\label{eq:SignalModel}
\end{equation}
where the matrix of the received samples is written as
\begin{equation*}
 	\mathbf{R} = \left[
   	\begin{matrix}
   		R_1(m) & \ldots & R_{N_\text{Rx}}(m)  \\
			\vdots & \ddots & \vdots \\
			R_1(m+l-1) & \ldots & R_{N_\text{Rx}}(m+l-1)
   	\end{matrix}
 	\right],
\end{equation*}
$\mathbf{G}$ is the $l \times N_{\text{Tx}}$ MIMO code generator matrix,
\begin{equation*}
 	\mathbf{H} = \left[
   	\begin{matrix}
   		H_{1,1}(m) & \ldots & H_{1,N_\text{Rx}}(m)  \\
			\vdots & \ddots & \vdots \\
			H_{1,1}(m+l-1) & \ldots & H_{1,N_\text{Rx}}(m+l-1)
   	\end{matrix}
 	\right],
\end{equation*}
is the $N_{\text{Tx}} \times N_{\text{Rx}}$ channel frequency response matrix that is assumed to be constant over the $l$ consecutive channel usage instants, and
\begin{equation*}
 	\mathbf{W} = \left[
   	\begin{matrix}
   		W_1(m) & \ldots & W_{N_\text{Rx}}(m)  \\
			\vdots & \ddots & \vdots \\
			W_1(m+l-1) & \ldots & W_{N_\text{Rx}}(m+l-1)
   	\end{matrix}
 	\right]
\end{equation*}
is the additive noise matrix.

\subsection{The Rayleigh Fading Channel Model}
The Rayleigh distributed complex channel coefficients were composed of independent Gaussian $\mathcal{N}(0,1/2)$ distributed real and imaginary parts i.e. their expected energy was one. The links between all $\text{Tx} - \text{Rx}$ antenna pairs were independent i.e. there was no correlation between the coefficients $H_{ij}(m)$.

A new channel was generated at the beginning of each MIMO code word. In the case of $3 \times N_\text{Rx}$ and $4 \times N_\text{Rx}$ MIMO, this implied a constant channel during four successive MIMO symbols, while for the Alamouti code, the channel changed after every two symbols. This is as close to full interleaving as we can get while keeping the channel constant within MIMO code words. The final shuffling was done by a random interleaver, the length of which was several thousands bits; the purpose of this shuffling was to scatter the bits within a single MIMO word all along the FEC code word. Hence, the model can be described as a fully interleaved flat fading Rayleigh channel.
%
%
\section{The Soft Decision Flow in the Receiver}
The basic idea behind delivering the soft decisions from the MIMO decoder to the FEC decoder is depicted in Fig.~\ref{fig:SystemModel}. Essentially three components that are subjects of concern: the MIMO decoder, the demodulator and the FEC decoder. The final aim is to deliver all information to the FEC decoder that is necessary for it to be able to find the ML code word. The way to achieve this is discussed in the following subsections.

\subsection{The Bitwise LLR Metric}
There are basically two ways to obtain an optimum FEC decoder. The traditional way is ML decoding, which is able to find the most probable code word and then output the corresponding information word. The second way is to maximise the \emph{a posteriori} probabilities (APP) of the information word symbols by applying the renowned BCJR algorithm~\cite{Bahl}. Berrou \emph{et al}. in the seminal paper~\cite{Berrou} very cleverly introduced an iterative decoding method based on the BCJR algorithm in which information is passed between the constituent decoders in the form of LLR. Berrou's invention
led to amazingly good performance of the code, which was
aptly named the `turbo code'.

The bitwise LLR is the logarithm in which the likelihood ratio of the bit is one versus zero, i.e.
\begin{equation}
	{\lambda_q} = \ln \left(\frac{p(x_q=1)}{p(x_q=0)}\right).
	\label{eq:BitwiseLlr}
\end{equation}
It is as optimal as the probabilities themselves --– which can easily be solved from $\lambda_q$. The use of LLR directly in decoding instead of probabilities often leads to a simpler decoder --– without compromising optimality e.g. it is a straightforward way to show that the summation of branch metric values that are in the form of LLR in a Viterbi decoder lead to the ML decoding result.

In the general case, calculating the true LLR is not a very simple task. The probabilities in a multilevel modulation come from multiple constellation points, which make the optimum demodulation less attractive. However, excellent approximations can be obtained with relative ease~\cite{Raju},~\cite{Gu}. To shed light on the bitwise LLR, we first consider BPSK modulated transmission of symbol values $\sqrt{E_s}$ and $-\sqrt{E_s}$ over the additive Gaussian channel $\mathcal{N}(0,\sigma^2)$. With the received symbol $R_q$ the probability density functions (pdf) become
\begin{align}
 	&p\left(x_q=1\right) = \frac{1}{\sqrt{2\pi\sigma^2}} \:
 		e^{-\left(R_q-\sqrt{E_s}\right)^2/\left(2\sigma^2\right)}	\nonumber \\
 	&\text{and}																		\nonumber \\
 	&p\left(x_q=0\right) = \frac{1}{\sqrt{2\pi\sigma^2}} \:
 		e^{-\left(R_q+\sqrt{E_s}\right)^2/\left(2\sigma^2\right)},
   \label{eq:GaussianPdf}
\end{align}
from which we obtain by direct calculation
\begin{equation}
	{\lambda_q} = \frac{2\sqrt{E_s}}{\sigma^2} R_q.
	\label{eq:LlrBpsk}
\end{equation}

This result can be generalised to QAM constellations as well. Suppose we have separate bit mappings onto real and imaginary parts of the constellation points and the Gray mapping of the bit vectors is used. Then the approximation of LLR can be obtained by measuring $\text{dist}\left(R_q,S_\text{mid}\right)$, which is the distance from the received point to the midpoint between the constellation points that give the opposite bit decisions and calculating
\begin{equation}
	{\lambda_q} = \frac{d_\text{const}}{\sigma^2} \text{dist}\left(R_q,S_\text{mid}\right),
	\label{eq:LlrQam}
\end{equation}
where $d_\text{const}$ is the distance between two constellation points, and all distances are measured either in real or imaginary dimensions.

\begin{figure}
	\centering
	\includegraphics[width=1.8in]{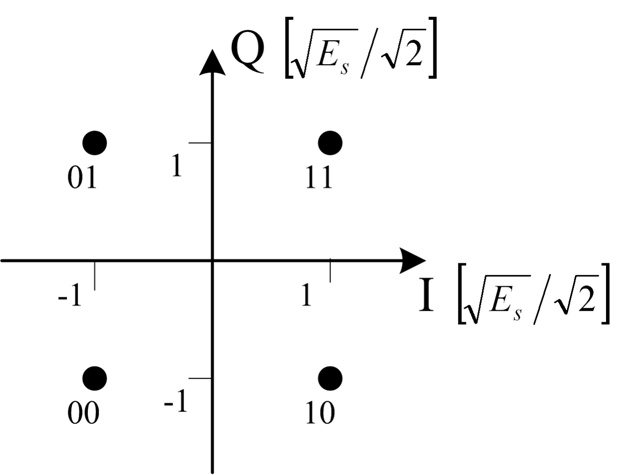}
 	\caption{QPSK constellation with Gray mapping, real and imaginary parts separated.}
	\label{fig:ConstellationQpsk}
\end{figure}

A simple example is the QPSK constellation depicted in Fig.~\ref{fig:ConstellationQpsk}, where the bitwise LLR values that are not approximations are obtained in the case $E_s=1$ as
\begin{equation}
	{\lambda_1} = \frac{\sqrt{2}\:x}{\sigma^2}, \qquad
	{\lambda_0} = \frac{\sqrt{2}\:y}{\sigma^2}.
	\label{eq:BitLlrBpsk}
\end{equation}

\begin{figure}
	\centering
	\includegraphics[width=3.2in]{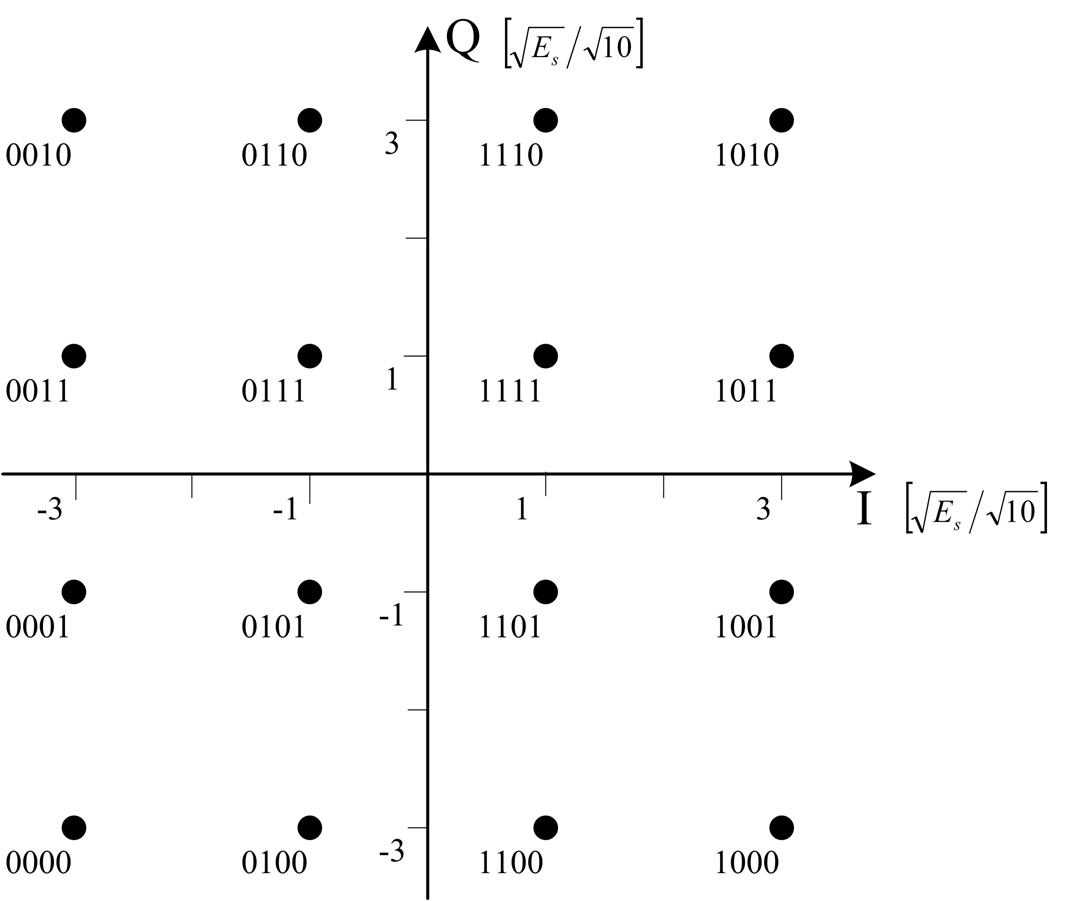}
 	\caption{16-QAM constellation with Gray mapping, real and imaginary parts separated.}
	\label{fig:Constellation16qam}
\end{figure}

A lucid example of the simplicity of the multilateral bitwise LLR metrics is the 16-QAM constellation depicted in Fig.~\ref{fig:Constellation16qam}, where the approximate LLR values in the case $E_s=1$ are calculated as
\begin{align}
	&{\lambda_3} \approx \frac{2x}{\sqrt{10}\:\sigma^2}, \qquad
	 {\lambda_2} \approx \frac{-2}{\sqrt{10}\:\sigma^2}\:
	             \left(\left|x\right|-\frac{2}{\sqrt{10}}\right),	\nonumber \\
	&{\lambda_1} \approx \frac{2y}{\sqrt{10}\:\sigma^2}, \qquad
	 {\lambda_0} \approx \frac{-2}{\sqrt{10}\:\sigma^2}\:
	             \left(\left|y\right|-\frac{2}{\sqrt{10}}\right).
	\label{eq:BitLlr16qam}
\end{align}

Hence, obtaining the approximate bitwise LLR can be even simpler than the hard decisions. Further, the approximations tend to be very close to the true values --– especially for the
small $\lambda_q$ values that are crucial to the FEC decoder's performance. The sub-optimality of \eqref{eq:BitLlr16qam}, then, has only a minor impact on the bit error rates (BER) at the FEC decoder output. According to our simulations, it is hard to discern any differences between the optimum and sub-optimum 16-QAM demodulations.

\subsection{The Vector LLR Metric}
\begin{figure}
	\centering
	\includegraphics[width=2.2in]{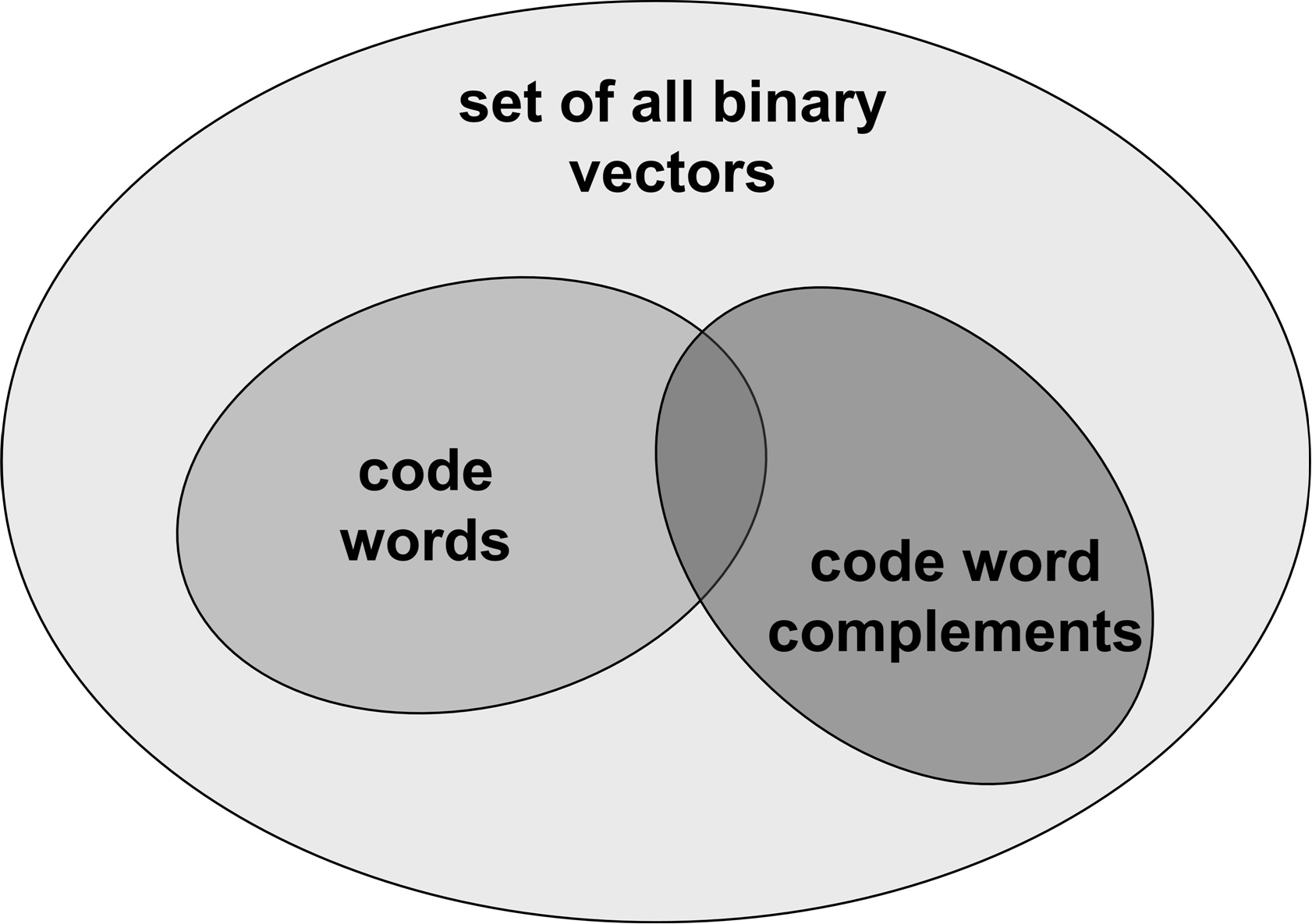}
 	\caption{The code words and code word complements presented as subsets of all binary vectors.}
	\label{fig:VectorSets}
\end{figure}

We aim to obtain the reliability metric that establishes a fair comparison between the code words --– or code vectors, which are essentially the same --– so that the ML decoder is able to pick the most probable code word. Given a set of binary vectors of length $L$ and the independent bit probabilities $p(x_q)$, the probability of the entire vector $p(\mathbf{x})=\prod_{q=0}^{L-1}p(x_q)$. Clearly $p(\mathbf{x}) > p(\mathbf{y})$ implies that $\mathbf{x}$ is more probable than $\mathbf{y}$. For the complement of $\mathbf{x}$, $p(\overline{\mathbf{x}})=\prod_{q=0}^{L-1}\left(1-p(x_q)\right)$, and the condition $p(\overline{\mathbf{x}}) < p(\overline{\mathbf{y}})$ holds if $\mathbf{x}$ is more probable. Therefore, the probability ratio
\begin{equation*}
	\frac{p(\mathbf{x})}{p(\overline{\mathbf{x}})} =
		\prod_{q=0}^{L-1} \frac{p(x_q)}{\left(1-p(x_q)\right)}
\end{equation*}
sorts the vectors so that the vector $\mathbf{x}$ is more probable than
the vector $\mathbf{y}$ whenever
\begin{equation*}
	\frac{p(\mathbf{x})}{p(\overline{\mathbf{x}})} >
		\frac{p(\mathbf{y})}{p(\overline{\mathbf{y}})}.
\end{equation*}
Using the logarithm does not change the order, since logarithm is a monotonically increasing function. Examining subsets instead of the complete set does not change the order either, which implies that the logarithm of the probability ratio sorts the words in the union of code words and code word complements, depicted in Fig.~\ref{fig:VectorSets}, so that most probable appears first.

Therefore, maximising the metric
\begin{equation}
	M_0(\Lambda) = \sum_{q=0}^{L-1}\text{sgn}\left(c_q\right)\lambda_q
	\label{eq:MetricLambda}
\end{equation}
among the code words $\left[c_0 \quad c_1 \quad \ldots \quad c_{L-1}\right]$ cranks out
the ML code word. The binary code word components in \eqref{eq:MetricLambda}, $c_q \in\left\{-1,1\right\}$ with the bit zero mapped onto $-1$.

\subsection{The Signal Energy and the Noise Variance}
Obtaining the $\lambda_q$ values and consequently the metric $M_0(\Lambda)$ seem very simple --– and it is, provided the noise variance $\sigma^2$ is same for every bit. However, in general it is not. The assumption that $\sigma^2$ is same at every Rx antenna is realistic, but what happens when the received raw signal is equalised? It is assumed that the phases and amplitudes of the constellation points are restored before reaching the demodulator, which essentially divides the received samples by the channel frequency responses $H_{ij}$. This implies that the additive noise samples also become divided by $H_{ij}$ and their variance by its square, i.e. by $H_{ij}^*H_{ij}$. Since $\sigma^2$ resides in the denominator of the expression of $\lambda_q$, each $\lambda_q$ becomes multiplied by $H_{ij}^*H_{ij}$ as a result of the equalisation. Therefore, if the additive noise variance is same at all Rx antennas throughout the FEC code block, then $\sigma^2$ in the expression of $\lambda_q$ can be set to unity provided we at same time multiply the $\lambda_q$ values by the energy of the channel coefficients that are related to the transmission of the given bit. In the MIMO case, this channel energy term is not just a single $H_{ij}^*H_{ij}$ but the double sum over all $ij$ combinations.

At this stage we have lost the probabilistic interpretation of $M_0(\Lambda)$, though this could be restored by finding a coefficient that would scale the sum of probabilities to the unity. This is unnecessary, however, since we are interested in the order of the $M_0(\Lambda)$, which is not affected when all metrics are multiplied by a same constant.

As an example, consider the transmission of a constellation point $S$ as the $m$th symbol in the transmission chain. The MRC combiner of the samples obtained from $N_\text{Rx}$ antennas calculates the estimate
\begin{equation}
	\hat{S}(m) = \frac
	   {\displaystyle{\sum_{j=1}^{N_{\text{Rx}}}R_j(m)H_{1j}^*(m)}}
		{\displaystyle{\sum_{j=1}^{N_{\text{Rx}}}H_{1j}^*(m)H_{1j}(m)}} 
	\label{eq:EstMrc}
\end{equation}
where the sum in the denominator is the channel energy term, by which the $\lambda_q$ values are to be multiplied.

As the second example, consider the Alamouti coded transmission from two Tx antennas
\begin{equation}
	\mathbf{R}=\mathbf{G}_2\mathbf{H}+\mathbf{W}
	\label{eq:SignalModelTx2}
\end{equation}
where the Alamouti code matrix
\begin{equation*}
 	\mathbf{G}_2 = \left[
   	\begin{matrix}
   		S_1(m)   & S_2(m) \\
			S_2^*(m) & S_1^*(m)
   	\end{matrix}
 	\right], \quad m=0,\,2,\,4,\,\ldots\;.
\end{equation*}
The Alamouti combiner calculates the estimates
\begin{align}
	&\hat{S}_1(m) = \frac
	   {\displaystyle{\sum_{j=1}^{N_{\text{Rx}}}R_j  (m)  H_{1j}^*(m) +
	                  \sum_{j=1}^{N_{\text{Rx}}}R_j^*(m+1)H_{2j}  (m)}}
		{\displaystyle{\sum_{j=1}^{N_{\text{Rx}}}\sum_{i=1}^{2}H_{ij}^*(m)H_{ij}(m)}},
		\nonumber \\
	&\hat{S}_2(m) = \frac
	   {\displaystyle{\sum_{j=1}^{N_{\text{Rx}}}R_j  (m)  H_{2j}^*(m) -
	                  \sum_{j=1}^{N_{\text{Rx}}}R_j^*(m+1)H_{1j}  (m)}}
		{\displaystyle{\sum_{j=1}^{N_{\text{Rx}}}\sum_{i=1}^{2}H_{ij}^*(m)H_{ij}(m)}},
	\label{eq:EstAlamouti}
\end{align}
where the double sum in the denominator i.e. the channel energy term is the same for both symbols.

To generalise, we state that maximising
\begin{equation}
	M(\Lambda) = \sum_{q=0}^{L-1}\text{sgn}\left(c_q\right)
	\sum_{j=1}^{N_{\text{Rx}}}\sum_{i=1}^{N_{\text{Tx}}}H_{ij}^*(m)H_{ij}(m)
	\lambda_{q\left|{\sigma^2=1}\right.} 
	\label{eq:MetricLambdaScale}
\end{equation}
leads to the ML decision on the transmitted code word. Note that although the double sum energy term varies in the transmission chain index $m$, a change can only occur at the beginning of a new MIMO code word e.g. the energy term is the same for every $\lambda_q$, the constellation points of which are transmitted within the same channel usage interval $m \; \ldots \; m+l-1$.

Note that the double sum channel energy term that appears in the expression of $M(\Lambda)$ has exactly the same form in the denominator of the Alamouti combiner, and it can be found in the MRC combiner as well by noticing there is only one Tx antenna to sum up. This coincidence might lead people to think that the optimum way to decode and combine the MIMO encoded signals involves the channel energy, which in fact it does --– we do not propose any other type of MIMO combiners in this paper. Further, the energy term we introduced in $M(\Lambda)$ did not come from the Alamouti combiner, but directly resulted from restoring the amplitude and phase of the transmitted symbol by dividing the received samples by $H_{ij}$. Therefore, the optimality of $M(\Lambda)$ is not restricted to the cases we present here.
%
%
\section{The Orthogonal Rate 3/4 MIMO Codes}
Tarokh \emph{et al}. called the rate 3/4 codes for three and four Tx antennas that they had discovered `sporadic'~\cite{TarokhTIT}. At that time, of course, they were sporadic in the sense that there were no rules on how to construct similar codes beyond four Tx antennas. Later, Liang discovered there rules~\cite{Liang}, which state that e.g. for five and six Tx antennas there exist rate 2/3 codes. Further, the code generators of Tarokh \emph{et al}. are not unique and the forms presented in \cite{Liang} are more convenient for decoding since the generator matrices contain only constellation points and their complex conjugates. However, we did not opt for either of the generators above mentioned, but selected yet another constructions presented by Giannakis \emph{et al}.~\cite{Giannakis}. The generators of  are essentially the same as those in \cite{Liang} with a few columns interchanged. The reason for using the generators of \cite{Giannakis} is simple: having the constellation points and their conjugates well ordered in the generator matrix makes the decoding processes more straightforward, and it ultimately satisfies important optimality criteria.

\subsection{The Code for Three Tx antennas}
\label{sec:IVA}
The generator matrix of the code~\cite{Giannakis} is
\begin{equation}
 	\mathbf{G}_3 = \left[
   	\begin{matrix}
   		 S_1   &  S_2   & S_3   \\
			-S_2^* &  S_1   & 0     \\
			-S_3^* &  0     & S_1^* \\
			 0     & -S_3^* & S_2^*
   	\end{matrix}
 	\right],
   \label{eq:Tx3GenMatrix}
\end{equation}
where $[S_1 \; S_2 \; S_3]$ are the three information bearing constellation points that are transmitted in the four consecutive instants.

The signal received by the MIMO decoder is modelled as
\begin{equation}
	\mathbf{R}=\mathbf{G}_3\mathbf{H}+\mathbf{W}
\label{eq:Tx3SignalModel}
\end{equation}
where
\begin{equation*}
 	\mathbf{R} = \left[
   	\begin{matrix}
   		R_1(m)   & \ldots & R_{N_\text{Rx}}(m)   \\
   		R_1(m+1) & \ldots & R_{N_\text{Rx}}(m+1) \\
   		R_1(m+2) & \ldots & R_{N_\text{Rx}}(m+2) \\
			R_1(m+3) & \ldots & R_{N_\text{Rx}}(m+3)
   	\end{matrix}
 	\right],
   \label{eq:Tx3ReceivedMatrix}
\end{equation*}
$\mathbf{G}$ is the $l \times N_{\text{Tx}}$ MIMO code generator matrix,
\begin{equation*}
 	\mathbf{H} = \left[
   	\begin{matrix}
   		H_{1,1}(m) & \ldots & H_{1,N_\text{Rx}}(m) \\
   		H_{2,1}(m) & \ldots & H_{2,N_\text{Rx}}(m) \\
			H_{3,1}(m) & \ldots & H_{3,N_\text{Rx}}(m)
   	\end{matrix}
 	\right],
   \label{eq:Tx3ChannelResponseMatrix}
\end{equation*}
conforming to the assumption $H_{ij}(m)\;\ldots\;H_{ij}(m+3)$.

The dimensions and indexing of $\mathbf{W}$ are identical to those of $\mathbf{R}$. On the receiver side, the minimum variance unbiased estimators of the transmitted constellation points are calculated in a fashion that resembles the Alamouti combiner. The equations derived in Appendix~\ref{sec:AppendixA} are given in~(\ref{eq:EstTx3Orthogonal}),
\begin{figure*} 
\begin{align}
	&\hat{S}_1(m) = \frac
	   {\displaystyle{\sum_{j=1}^{N_{\text{Rx}}}R_j  (m)  H_{1j}^* +
	                  \sum_{j=1}^{N_{\text{Rx}}}R_j^*(m+1)H_{2j}   +
	                  \sum_{j=1}^{N_{\text{Rx}}}R_j^*(m+2)H_{3j}     }}
		{\displaystyle{\sum_{j=1}^{N_{\text{Rx}}}\sum_{i=1}^{3}H_{ij}^*H_{ij}}},
		\nonumber \\
	&\hat{S}_2(m) = \frac
	   {\displaystyle{\sum_{j=1}^{N_{\text{Rx}}}R_j  (m)  H_{2j}^* -
	                  \sum_{j=1}^{N_{\text{Rx}}}R_j^*(m+1)H_{1j}   +
	                  \sum_{j=1}^{N_{\text{Rx}}}R_j^*(m+3)H_{3j}     }}
		{\displaystyle{\sum_{j=1}^{N_{\text{Rx}}}\sum_{i=1}^{3}H_{ij}^*H_{ij}}},
		\nonumber \\
	&\hat{S}_3(m) = \frac
	   {\displaystyle{\sum_{j=1}^{N_{\text{Rx}}}R_j  (m)  H_{3j}^* -
	                  \sum_{j=1}^{N_{\text{Rx}}}R_j^*(m+2)H_{1j}   -
	                  \sum_{j=1}^{N_{\text{Rx}}}R_j^*(m+3)H_{2j}     }}
		{\displaystyle{\sum_{j=1}^{N_{\text{Rx}}}\sum_{i=1}^{3}H_{ij}^*H_{ij}}}
	\label{eq:EstTx3Orthogonal}
\end{align}
\hrulefill
\end{figure*}
where for brevity, the transmission chain index $m$ is only marked onto $R_j$.

The combiner of the orthogonal rate 3/4 code for three Tx antennas bears an obvious resemblance to both MRC and Alamouti combiners, and it can be considered an extension of them. As shown in  Appendix~\ref{sec:AppendixB}, it even satisfies the MRC limit, thus attaining the highest achievable SNR in the combined estimates.

\subsection{The Code for Four Tx antennas}
The construction is very similar with that in Section~\ref{sec:IVA}.
The generator matrix of the code~\cite{Giannakis} is
\begin{equation}
 	\mathbf{G}_3 = \left[
   	\begin{matrix}
   		 S_1   &  S_2   &  S_3   &  0   \\
			-S_2^* &  S_1   &  0     &  S_3 \\
			-S_3^* &  0     &  S_1^* & -S_2 \\
			 0     & -S_3^* &  S_2^* &  S_1
   	\end{matrix}
 	\right],
   \label{eq:Tx4GenMatrix}
\end{equation}
where $[S_1 \; S_2 \; S_3]$ are the three information bearing constellation points that are transmitted in the four consecutive instants.

The signal received by the MIMO decoder is modelled as
\begin{equation}
	\mathbf{R}=\mathbf{G}_4\mathbf{H}+\mathbf{W}
\label{eq:Tx4SignalModel}
\end{equation}
wherethe $\mathbf{R}$ and $\mathbf{W}$ are the same as in the previous section and
\begin{equation*}
 	\mathbf{H} = \left[
   	\begin{matrix}
   		H_{1,1}(m) & \ldots & H_{1,N_\text{Rx}}(m) \\
   		H_{2,1}(m) & \ldots & H_{2,N_\text{Rx}}(m) \\
   		H_{3,1}(m) & \ldots & H_{3,N_\text{Rx}}(m) \\
			H_{4,1}(m) & \ldots & H_{4,N_\text{Rx}}(m)
   	\end{matrix}
 	\right],
   \label{eq:Tx4ChannelResponseMatrix}
\end{equation*}
conforming to the assumption $H_{ij}(m)\;\ldots\;H_{ij}(m+3)$.

The minimum variance unbiased estimators of the transmitted constellation points are given in~(\ref{eq:EstTx4Orthogonal}),
\begin{figure*} 
\begin{align}
	&\hat{S}_1(m) = \frac
	   {\displaystyle{\sum_{j=1}^{N_{\text{Rx}}}R_j  (m)  H_{1j}^* +
	                  \sum_{j=1}^{N_{\text{Rx}}}R_j^*(m+1)H_{2j}   +
	                  \sum_{j=1}^{N_{\text{Rx}}}R_j^*(m+2)H_{3j}   +
	                  \sum_{j=1}^{N_{\text{Rx}}}R_j  (m+3)H_{4j}^*   }}
		{\displaystyle{\sum_{j=1}^{N_{\text{Rx}}}\sum_{i=1}^{4}H_{ij}^*H_{ij}}},
		\nonumber \\
	&\hat{S}_2(m) = \frac
	   {\displaystyle{\sum_{j=1}^{N_{\text{Rx}}}R_j  (m)  H_{2j}^* -
	                  \sum_{j=1}^{N_{\text{Rx}}}R_j^*(m+1)H_{1j}   -
	                  \sum_{j=1}^{N_{\text{Rx}}}R_j  (m+2)H_{4j}^* +
	                  \sum_{j=1}^{N_{\text{Rx}}}R_j^*(m+3)H_{3j}     }}
		{\displaystyle{\sum_{j=1}^{N_{\text{Rx}}}\sum_{i=1}^{4}H_{ij}^*H_{ij}}},
		\nonumber \\
	&\hat{S}_3(m) = \frac
	   {\displaystyle{\sum_{j=1}^{N_{\text{Rx}}}R_j  (m)  H_{3j}^* +
	                  \sum_{j=1}^{N_{\text{Rx}}}R_j  (m+1)H_{4j}^* -
	                  \sum_{j=1}^{N_{\text{Rx}}}R_j^*(m+2)H_{1j}   -
	                  \sum_{j=1}^{N_{\text{Rx}}}R_j^*(m+3)H_{2j}     }}
		{\displaystyle{\sum_{j=1}^{N_{\text{Rx}}}\sum_{i=1}^{4}H_{ij}^*H_{ij}}}
	\label{eq:EstTx4Orthogonal}
\end{align}
\hrulefill
\end{figure*}
where for brevity, the transmission chain index $m$ is only marked onto $R_j$. The derivation that is essentially similar to the one with three Tx antennas is omitted.

Again the similar results are obtained. The combiner of the orthogonal rate 3/4 code --– this time for four Tx antennas –-- resemblances the combiners for one, two and three Tx antennas and can be considered an extension of them. The combiner also satisfies the MRC limit, the proof of which is essentially similar to that where there were three Tx antennas is omitted.
%
%
\section{Performance Comparison of the Orthogonal MIMO Codes}

\subsection{The Demodulator Output}
\begin{figure}
	\centering
	\includegraphics[width=3.4in]{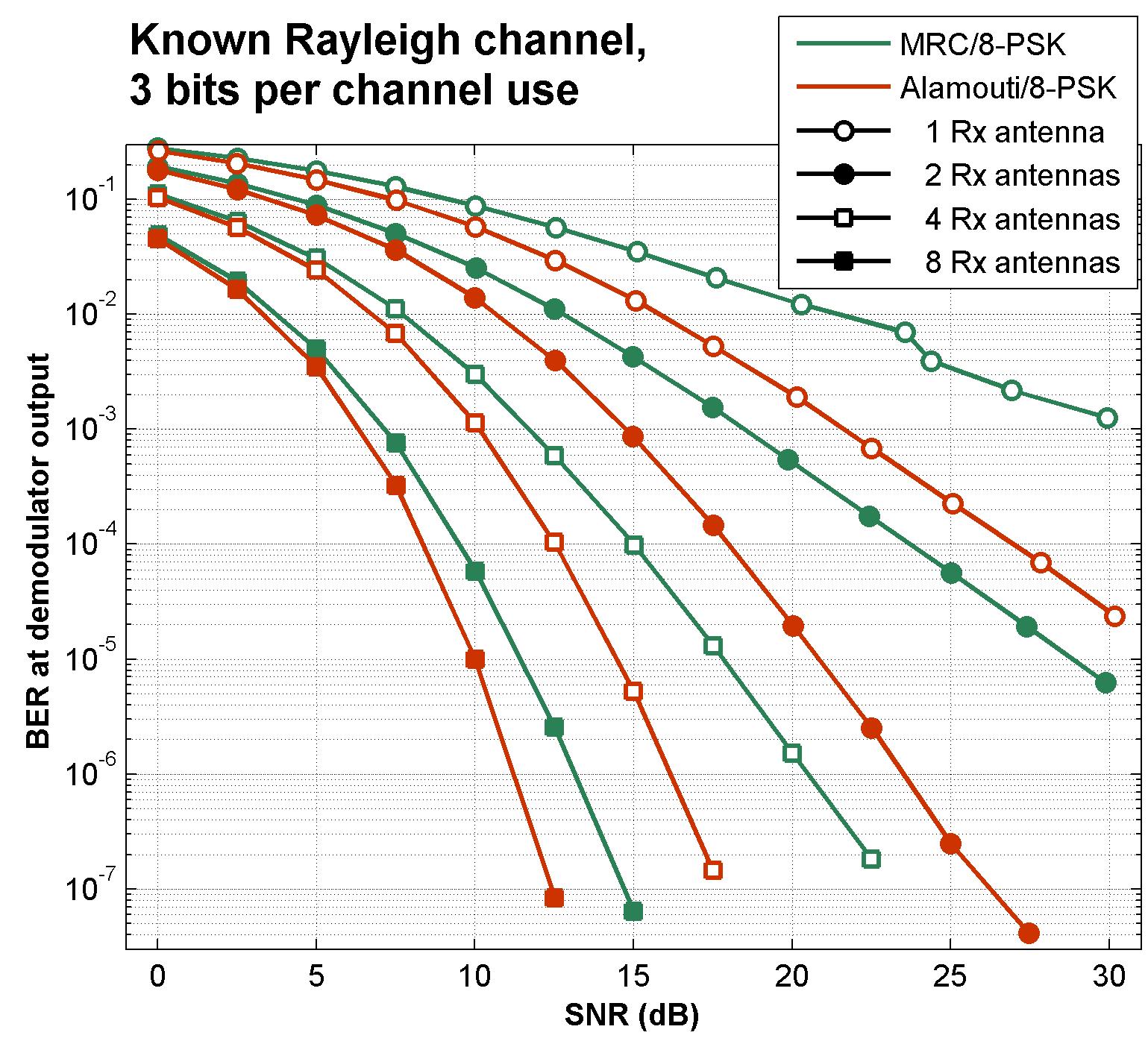}
 	\caption{Demodulator output BER of MRC and the Alamouti code in 8-PSK constellation mapping.}
	\label{fig:DemodBerTx1Tx2}
\end{figure}

When studying the performance without the FEC code, the abscissa of the curves is the average SNR received by one antenna and is not to be mixed with the compound SNR. Examining the BER vs. SNR curves at the demodulator output is tantamount to studying bitwise hard decisions, a topic which has already been addressed in the literature. However, we present some curves that are essential to understanding the diversity gain that is attainable by concatenation of the MIMO and FEC codes. Proper yardsticks for comparisons are found in the seminal papers of Alamouti and Tarokh \emph{et al}.~\cite{Alamouti},~\cite{TarokhJSac}. In Fig.~\ref{fig:DemodBerTx1Tx2}, the comparison of MRC and the Alamouti code is done by extending the results in \cite{Alamouti} to 8-PSK. The 3 dB shift is worth noting between MIMO schemes with an identical diversity order, which is visible both in our and Alamouti's curves. This can be intuitively explained by noting that even though the energy and noise terms of the MRC and Alamouti combiners look as if they were the same, the first Alamouti combiner equation removes the second transmitted symbol whereas the second one removes the first symbol. That is to say, with only one symbol energy the Alamouti combiner loses 3 dB in comparison with MRC.

Fig. 6 is an extension of the results in \cite{TarokhJSac} to multiple Rx antennas. In comparing the $1\times4$ and $4\times1$ MIMO schemes, we see that the BER curves are parallel, and it is straightforward to expand the MIMO combiner equations and show that they attain the same diversity order. However, the difference between the curves is 6 dB, which has an obvious explanation. Namely, the combiner cancels two parallel symbols, in addition to which the MIMO code rate 3/4 reduces the efficiency. Indeed, $10\log_{10}(1/3 \cdot 3/4) = -6\,\text{dB}$, though we cannot claim that this result is accurate due to different constellation mappings. The same difference can
also be seen between the $1\times8$ and $4\times2$ MIMO curves.

\begin{figure}
	\centering
	\includegraphics[width=3.4in]{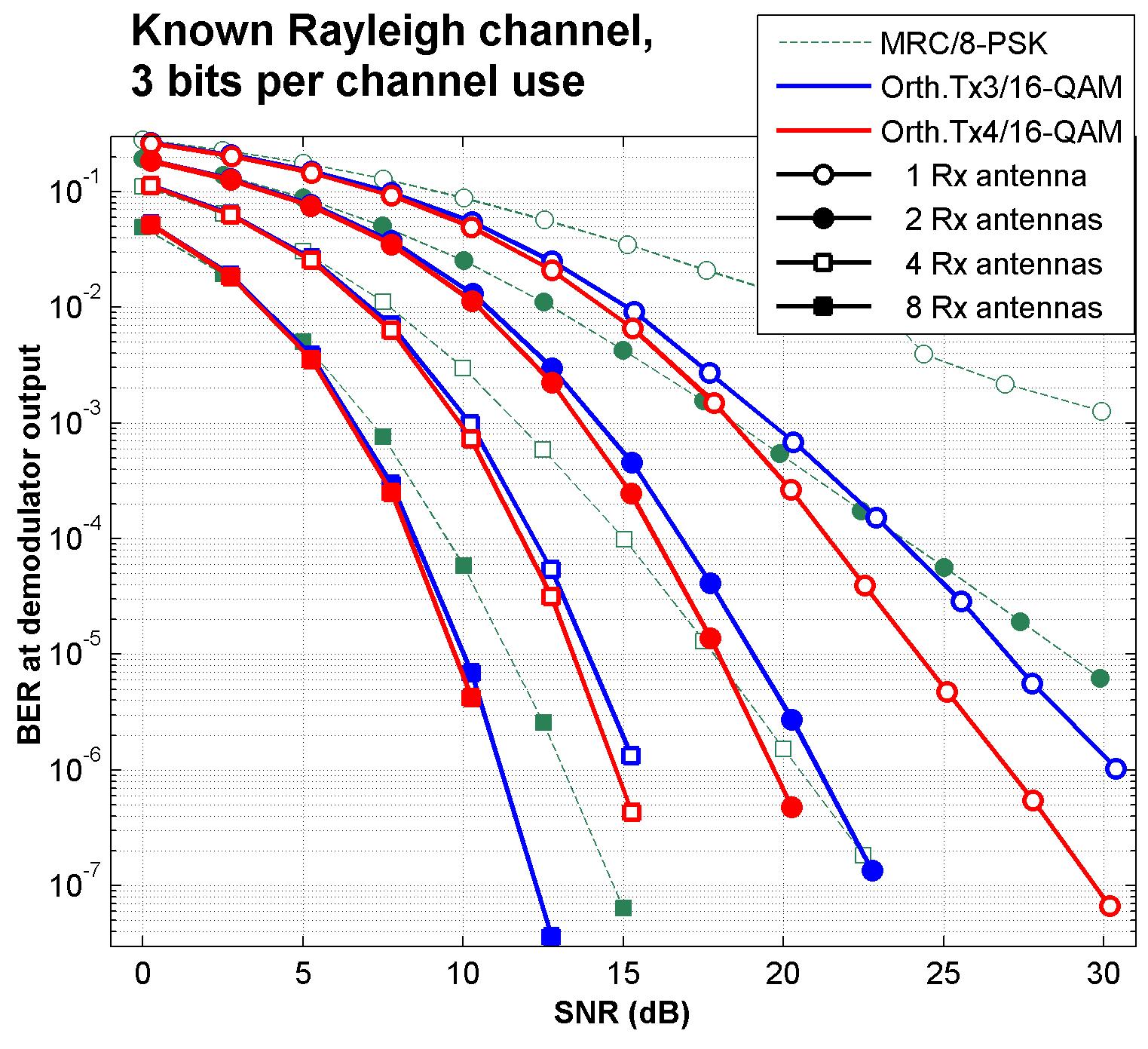}
 	\caption{Demodulator output BER of the rate 3/4 MIMO codes in 16-QAM constellation mapping.}
	\label{fig:DemodBerTx3Tx4}
\end{figure}

The diversity order of an $N_{\text{Tx}} \times N_{\text{Rx}}$ MIMO is the number of branches i.e. the connections from the Tx to the Rx antennas, which is the product $N_{\text{Tx}}N_{\text{Rx}}$. The $1 \times N_{\text{Rx}}$ MIMO is ideal in the sense that the expected received signal energy increases linearly in the diversity order. Adding more Tx antennas increases the diversity order but not the expected received signal energy, from which we can infer that an $N_{\text{Tx}} \times N_{\text{Rx}}$ MIMO receives $1/N_{\text{Tx}}$ of the energy of MRC that has the same diversity order.

To generalise, we state that for the combiner of any orthogonal complex MIMO code, the expected received symbol energy in comparison with MRC can be upper bounded as
\begin{equation}
	\Delta_{N_{\text{Tx}}} = 10\log_{10}(1/N_{\text{Tx}}) \quad [\text{dB}],
	\label{eq:CombinerUpperBound}
\end{equation}
where we of course assume the same diversity orders. The bound was seen to be met with equality in the case of two and four Tx antennas, meanwhile the case of three Tx antennas was slightly inferior.

To prove that the bound (\ref{eq:CombinerUpperBound}) is met with equality for any even $N_{\text{Tx}}$ and square generator matrix we note that in accordance with (\ref{eq:WangBound}), the maximum rate MIMO code has $N_{\text{Tx}}/2+1$ complex information bearing constellation points. If the combiner equations exist, each must cancel the energies of all other $N_{\text{Tx}}/2$ symbols leaving just one, from which we obtain the relative loss
\begin{equation}
	\frac{\left\lceil N_{\text{Tx}}/2 \right\rceil+1}{2\left\lceil N_{\text{Tx}}/2 \right\rceil}
	 \cdot \frac{1}{N_{\text{Tx}}/2+1} = \frac{1}{N_{\text{Tx}}}.
	\label{eq:RelativeLoss}
\end{equation}

An even $N_{\text{Tx}}$ and a square generator matrix form the ideal case, in comparison to which an odd $\Delta_{N_{\text{Tx}}}$ and a rectangular generator matrix are always inferior. As an example, the generator of the orthogonal code for three Tx antennas is a rectangular $4\times3$ matrix, and indeed, its combined signal energies remain slightly below the bound.

When the bound $\Delta_{N_{\text{Tx}}}$ is achieved, the orthogonal complex MIMO codes look as if they were optimal, which they aren't --- except for the Alamouti code. With more than two Tx antennas the number of information bearing constellation points remain lower than $\Delta_{N_{\text{Tx}}}$, and the energy is bounded in $\Delta_{N_{\text{Tx}}}$.

There are non-orthogonal codes that are able to outperform the orthogonal ones when the error rates at the MIMO decoder output are the yardstick. Damen \emph{et al}.~\cite{DamenNumT} reported that even the Alamouti code could be outperformed by a witty algebraic MIMO construction, which is a significant result. However, applying soft decisions and measuring the performance at the FEC decoder output is a different matter, and in this case the Alamouti code has yet to be beaten. With a larger number of Tx antennas, it seems feasible that the orthogonal complex MIMO codes could be outperformed --- even with the FEC code --- though the question of the soft decision MIMO decoder remains to be settled.

\subsection{The FEC decoder Output}
\begin{figure}
	\centering
	\includegraphics[width=3.4in]{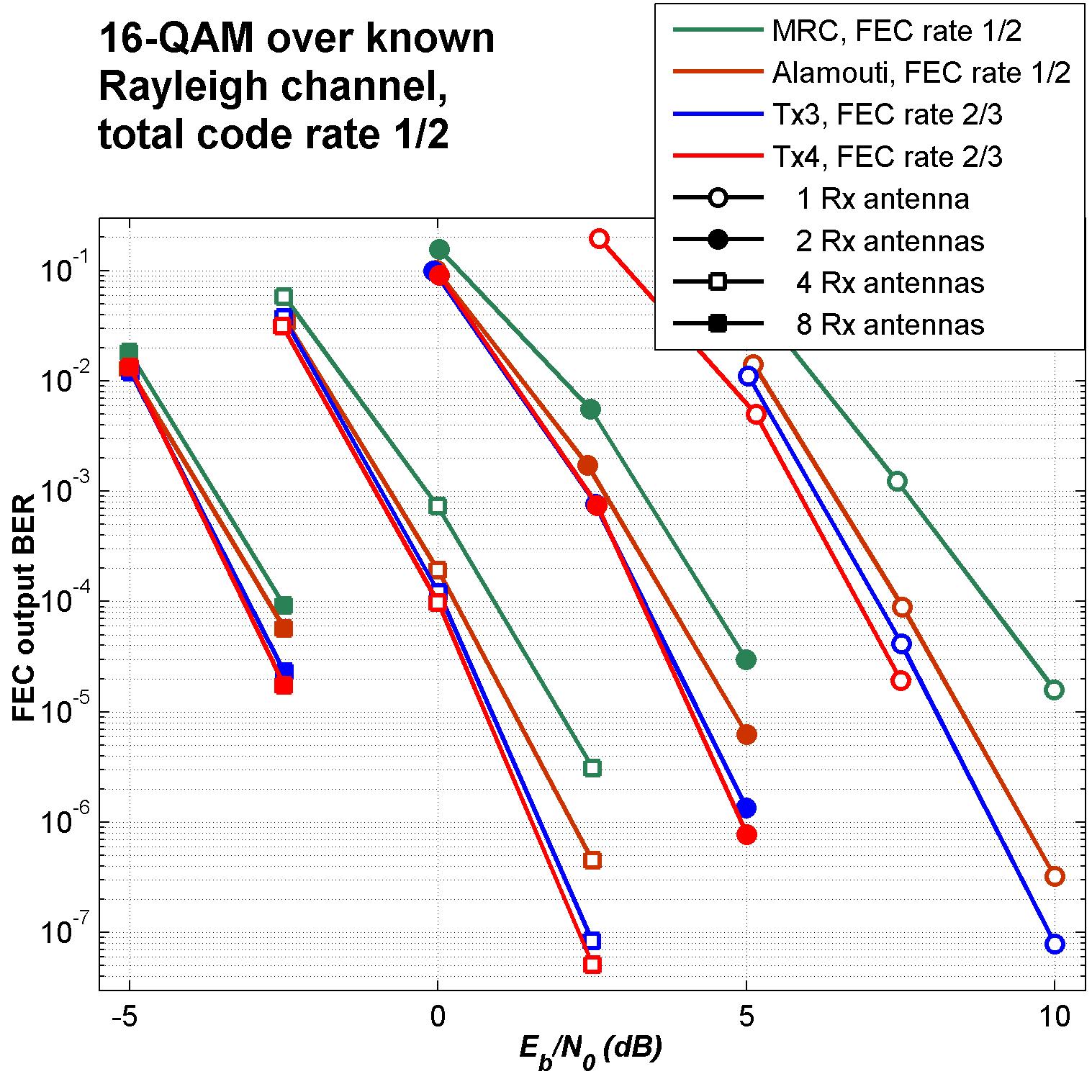}
 	\caption{Performances of the soft decision MIMO codes measured at the FEC decoder output.}
	\label{fig:FecBerSoftDec}
\end{figure}

The performance of the system with the FEC code is measured as BER versus $E_b/N_0$, where $E_b$ is the average information bit energy received by one antenna and $N_0$ is the single sided noise spectral density. The FEC code we used in simulations was a simple rate $1/2$, memory 6 convolutional code with the generators $133_8$ and $171_8$. In the case of rate $3/4$ MIMO codes, the total code rate was adjusted to $1/2$ by puncturing the convolutional code with the matrix 
$\left[ {}^1_1 \; {}^1_0 \right]$ to reach the rate of $2/3$. The simulation results with 16-QAM
over the fully interleaved flat Rayleigh channel are presented in Fig.~\ref{fig:FecBerSoftDec}. There we applied the full soft decision flow including the bitwise LLR demodulator as depicted in Fig.~\ref{fig:SystemModel}.

What is surprising in our results is that the performance is mainly dictated by the number of Rx antennas, while the number of Tx antennas and the MIMO code itself have only a minor impact on it. The explanation can be discerned in Fig.~\ref{fig:DemodBerTx3Tx4}, where the curves with the same number of Rx antennas are located in a same bundle at the demodulator output $\text{BER} \geq 10^2$. The Alamouti code curves in Fig.~\ref{fig:DemodBerTx1Tx2} closely fall into the same bundles, while MRC is slightly inferior. This is the raw error rate where the convolutional code starts to provide substantial coding gain, and where the differences between the MIMO schemes are determined --- or at least those differences we have the recourses to simulate. At extremely low BER, we obtain additional gain from the larger diversity order.

\subsection{Hard Decision vs. Soft Decision}
\begin{figure}
	\centering
	\includegraphics[width=3.4in]{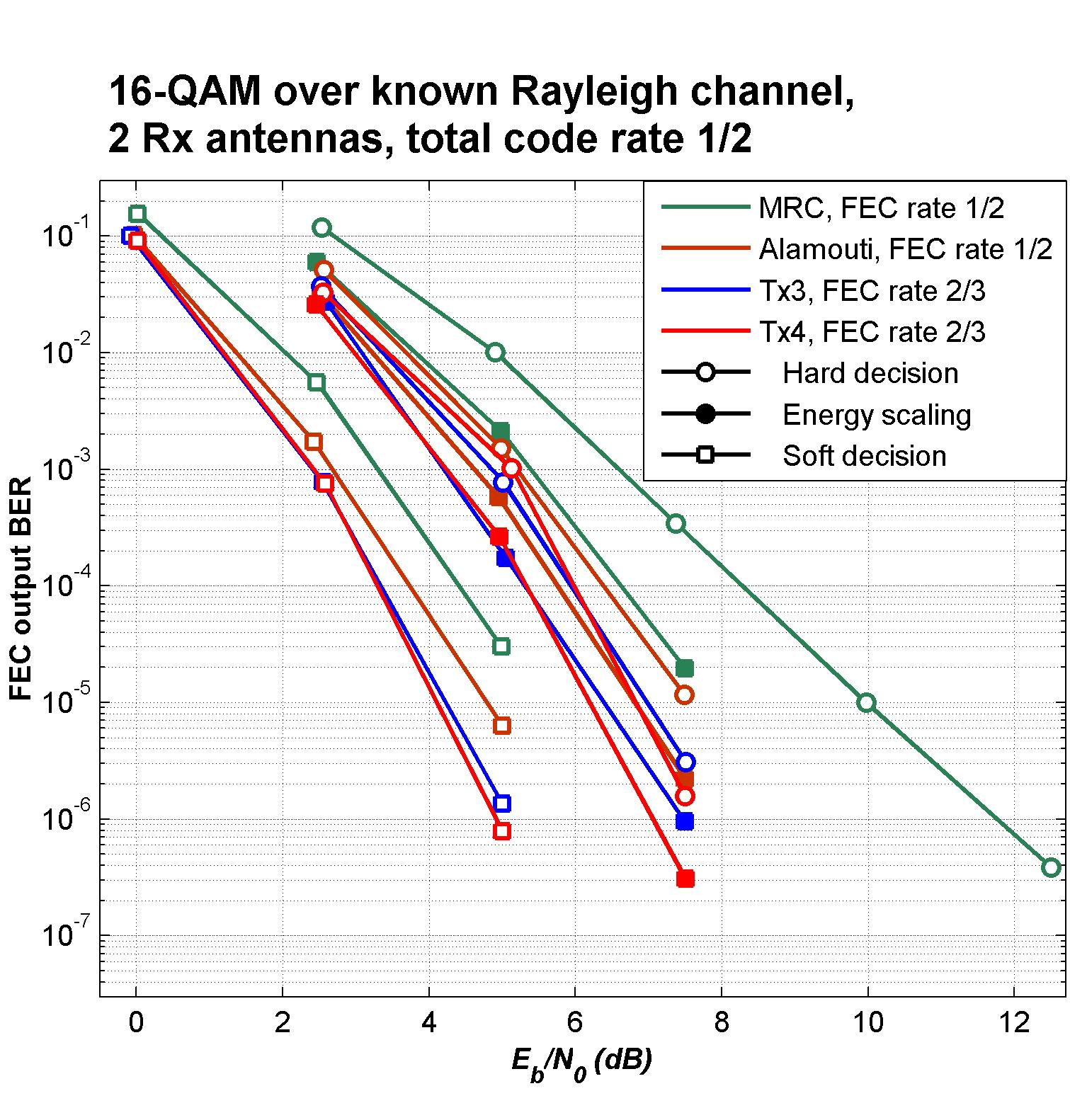}
 	\caption{Comparison of the hard and soft decision performances at the FEC decoder output.}
	\label{fig:FecBerHardSoftDec}
\end{figure}

Although the curves in Fig.~\ref{fig:FecBerSoftDec} manifest reliable transmission at remarkably low $E_b/N_0$, they do not tell exactly how much of a gain the soft decision MIMO coding with bitwise LLR demodulation has provided. Thus, we select two Rx antennas as a representative case to present the comparisons between the hard and soft decisions in Fig.~\ref{fig:FecBerHardSoftDec}. In the third, the energy scaling curve, the MIMO decoder made hard decisions, but the bit values were later mapped onto the set $\left\{1,-1\right\}$ and multiplied by the channel energy. The gain from the soft decisions was nearly $3 \: \text{dB}$ and it was even larger in MRC. This is a very significant result. The energy scaling yields the gain of $0.5$ to $1.0$~dB, and the rest comes from the soft decision MIMO combiners and bitwise LLR demodulator.

\subsection{Performance in OFDM over MIMO UWB Channel}
\begin{figure}
	\centering
	\includegraphics[width=3.4in]{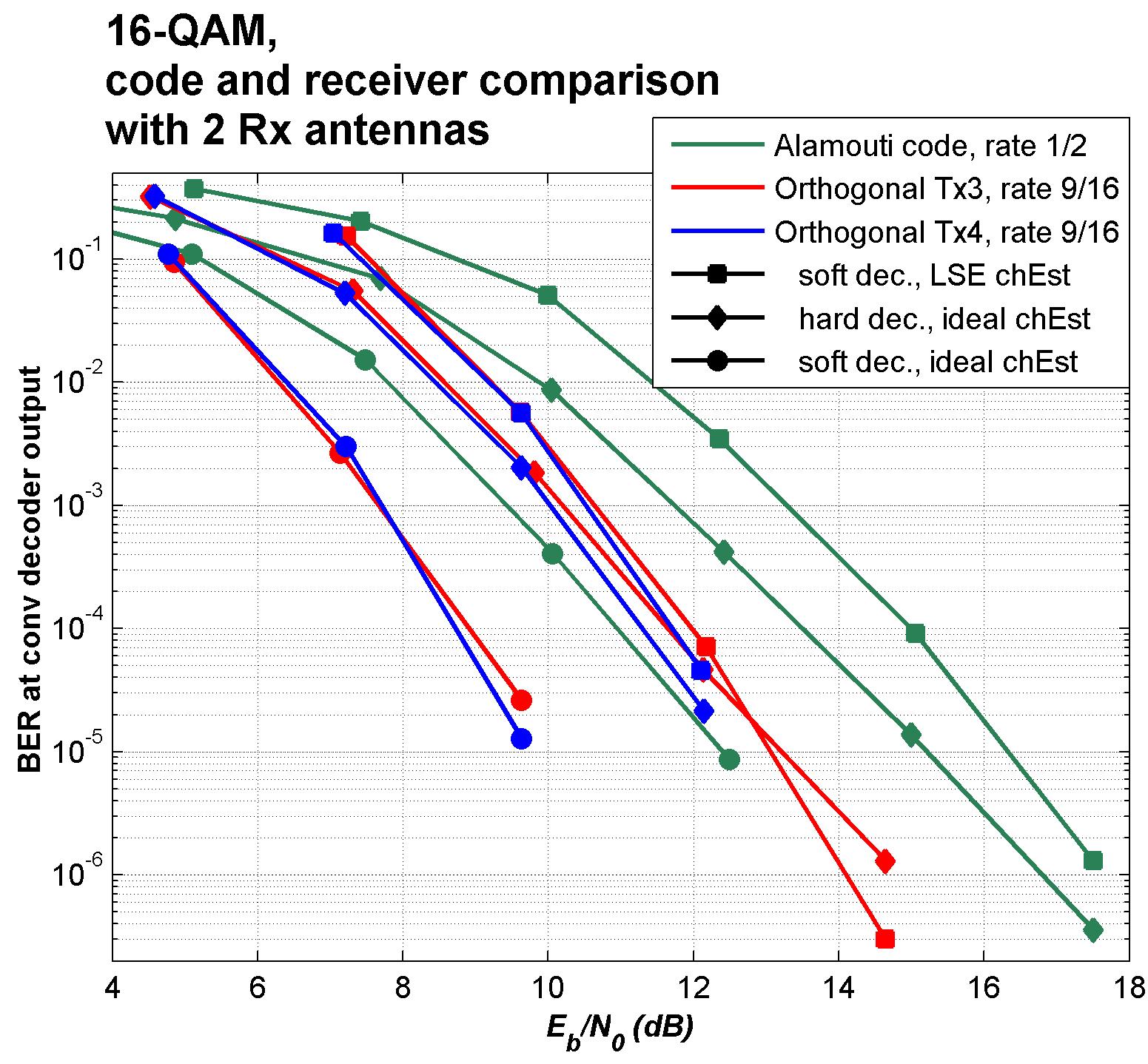}
 	\caption{Comparison of the orthogonal code performances in UWB OFDM transmission.}
	\label{fig:FecBerUwbOfdm}
\end{figure}

The fully interleaved Rayleigh channel is an idealisation that is rarely if ever encountered in practice. To gain some understanding of the real value of the soft decision MIMO receiver and orthogonal MIMO codes, we present some simulation results over a UWB channel in Fig.~\ref{fig:FecBerUwbOfdm}. The simulated model was OFDM and compatible with the ECMA-
368 standard~\cite{ECMA}. The FFT size was 128 and the OFDM symbols consisted of 100 data tones, 12 pilot tones 10 guard tones and 6 zero tones. Further, there were training symbols in front of each OFDM frame. This is the explanation for the higher $E_b/N_0$ values in Fig.~\ref{fig:FecBerUwbOfdm} compared with Fig.~\ref{fig:FecBerHardSoftDec} --- the UWB channel does not differ so much from the fully interleaved Rayleigh. The channel that is described in detail in~\cite{Batra} was CM1, with a frequency band from 3168 to 3696 MHz; here the correlation between the antenna links was generated by a simplified model reported in~\cite{Adeane}. The channel was kept constant during the OFDM frames and changed at the beginning of a new frame.

We were forced to make one small amendment to ECMA-368 when constructing the simulation model. Viz. the 16-QAM mapping in the standard did not suit our bitwise LLR demodulator, and it was changed to the Gray mapping depicted in Fig.~\ref{fig:Constellation16qam}.

The FEC code given in ECMA-368 is a memory 6 convolutional code with the generators $133_8$ and $145_8$. It has exactly the same weight enumerator as the FEC code used in the simulations over the Rayleigh channel, which makes their performances equal --- unless the codes are punctured. There was no proper rate $2/3$ puncture pattern for the ECMA-368 code, which is why we applied the rate $3/4$ puncturing in connection with the rate $3/4$ MIMO codes. The resulting total rate $9/16$ is close enough to $1/2$ to justify nearly the same comparison.

In OFDM transmission over the UWB channel, the rate $3/4$ orthogonal codes outweighed the Alamouti code by 2 to 2.5 dB. Such a difference can be explained by noting that there were subtones in relatively deep fade that affected performance more than the randomly scattered Rayleigh fading antenna links. This in turn means that the gain from a larger diversity order becomes substantial. In Fig.~\ref{fig:FecBerUwbOfdm}, we also included curves that demonstrate the impact of channel estimation. Our combiners are optimal only when the channel is known, and there is a risk that inaccurate estimates deteriorate the performance. In the current example, the deterioration had the same order of magnitude as the improvement from the soft decisions.
%
%
\section{Summary}
We have extended the soft decision decoding and signal combining algorithms of MRC and the Alamouti code to the orthogonal MIMO codes for three and four Tx antennas. The complexities of the novel combiners do not essentially exceed that of the Alamouti combiner, making it very attractive in terms of implementation. With the soft decision MIMO combiners, the FEC decoders attain the same error rates with 2 to 3 dB smaller data bit energy than with the hard decision MIMO. The novel combiners require a close to constant channel over four instants, implying that they are better suited for low mobility environments.

The soft decision MIMO combiners provide the largest gain when linked to a bitwise soft decision demodulator. Extracting the bitwise LLR values from the combiner output symbols was shown to be a very simple task --- even though a multilevel constellation is a matter of concern. The bitwise LLR values weighted by the pertaining channel tap energies encompass the sufficient statistic for the FEC decoder so as to be able to find the ML code word.

Although the diversity order of the orthogonal MIMO codes for 1 to 4 Tx antennas depends only on the number of Tx -- Rx antenna links, this doesn't mean that they perform identically. One Tx antenna is the ideal because $10\log_{10}\left(1/N_\text{Tx}\right)$ dB of received energy is lost when new Tx antennas are added and the number of Rx antennas is reduced so that the diversity order remains the same. A combiner of an orthogonal complex MIMO code for an even numbers of Tx antennas can be able to collect all available energy. Despite that the coding is not optimum, since the number of information bearing symbols remains below $N_\text{Tx}$.

\appendices
%
\section{The Orthogonal Code for three Tx Antennas: Derivation of the Decoding and Combining Rules}
\label{sec:AppendixA}
For brevity of notation, we omit the transmission chain index $m$ from $S_i$ and $H_{ij}$, which are not subject to change within MIMO code words. The encoding rule $\mathbf{R}=\mathbf{G}_3\mathbf{H}+\mathbf{W}$ written out into the set of equations is thus
\begin{align*}
	&R_j(m)   =  S_1    H_{1j} + S_2    H_{2j} + S_3    H_{3j} + W_j(m),   \\
	&R_j(m+1) = -S_2^*  H_{1j} + S_1^*  H_{2j} + 0\cdot H_{3j} + W_j(m+1), \\
	&R_j(m+2) = -S_3^*  H_{1j} + 0\cdot H_{2j} + S_1^*  H_{3j} + W_j(m+2), \\
	&R_j(m+3) =  0\cdot H_{1j} - S_3^*  H_{2j} + S_2^*  H_{3j} + W_j(m+3).
\end{align*}

In decoding we consider first the case with one Rx antenna. To obtain the estimate of $S_1$ at the \emph{j}th Rx antenna we calculate the auxiliary variable
\begin{equation*}
	\breve{S}_{1j} = R_j(m)H_{1j}^* + R_j^*(m+1)H_{2j} + R_j^*(m+2)H_{3j},
\end{equation*}
which expands to
\begin{align*}
	\breve{S}_{1j}
	&= \left[ S_1 H_{1j} + S_2 H_{2j} + S_3 H_{3j} + W_j(m)\right]H_{1j}^*    \\
	&\quad+ \left[-S_2 H_{1j}^* + S_1 H_{2j}^*     + W_j^*(m+1)\right]H_{2j}  \\
	&\quad+ \left[-S_3 H_{1j}^* + S_1 H_{3j}^*     + W_j^*(m+2)\right]H_{3j}  \\
	&= \left[H_{1j}^*H_{1j} + H_{2j}^*H_{2j} + H_{3j}^*H_{3j}\right]S_1       \\
	&\quad + W_j(m)H_{1j}^* + W_j^*(m+1)H_{2j}+ W_j^*(m+2)H_{3j}.
\end{align*}

To estimate $S_2$, we calculate
\begin{equation*}
	\breve{S}_{2j} = R_j(m)H_{2j}^* - R_j^*(m+1)H_{1j} + R_j^*(m+3)H_{3j},
\end{equation*}
and expand it to
\begin{align*}
	\breve{S}_{2j}
	&= \left[ S_1 H_{1j} + S_2 H_{2j} + S_3 H_{3j} + W_j(m)\right]H_{2j}^*  \\
	&\quad- \left[-S_2 H_{1j}^* + S_1 H_{2j}^* + W_j^*(m+1)\right]H_{1j}    \\
	&\quad+ \left[-S_3 H_{2j}^* + S_2 H_{3j}^* + W_j^*(m+3)\right]H_{3j}    \\
	&= \left[H_{1j}^*H_{1j} + H_{2j}^*H_{2j} + H_{3j}^*H_{3j}\right]S_2     \\
	&\quad - W_j^*(m+1)H_{1j} + W_j(m)H_{2j}^*+ W_j^*(m+3)H_{3j}.
\end{align*}

The third auxiliary variable we choose as
\begin{equation*}
	\breve{S}_{3j} = R_j(m)H_{3j}^* - R_j^*(m+2)H_{1j} - R_j^*(m+3)H_{2j},
\end{equation*}
and expand it to
\begin{align*}
	\breve{S}_{3j}
	&= \left[ S_1 H_{1j} + S_2 H_{2j} + S_3 H_{3j} + W_j(m)\right]H_{3j}^*  \\
	&\quad- \left[-S_3 H_{1j}^* + S_1 H_{3j}^* + W_j^*(m+2)\right]H_{1j}    \\
	&\quad- \left[-S_3 H_{2j}^* + S_2 H_{3j}^* + W_j^*(m+3)\right]H_{2j},   \\
	&= \left[H_{1j}^*H_{1j} + H_{2j}^*H_{2j} + H_{3j}^*H_{3j}\right]S_3     \\
	&\quad - W_j^*(m+2)H_{1j} - W_j(m+3)H_{2j}^*+ W_j^*(m)H_{3j}.
\end{align*}

Consequently, the minimum variance unbiased estimators of the transmitted constellation points for one Rx antenna are
\begin{align*}
	\\
	\hat{S}_{1j}
	&= \frac{R_j(m)H_{1j}^* + R_j^*(m+1)H_{2j} + R_j^*(m+2)H_{3j}} 
	        {\displaystyle{\sum_{i=1}^3 H_{ij}^*H_{ij}}},
	\\ \\
	\hat{S}_{2j}
	&= \frac{R_j(m)H_{2j}^* - R_j^*(m+1)H_{1j} + R_j^*(m+3)H_{3j}} 
	        {\displaystyle{\sum_{i=1}^3 H_{ij}^*H_{ij}}},
	\\ \\
	\hat{S}_{3j}
	&= \frac{R_j(m)H_{3j}^* - R_j^*(m+2)H_{1j} - R_j^*(m+3)H_{2j}} 
	        {\displaystyle{\sum_{i=1}^3 H_{ij}^*H_{ij}}}.
	\\
\end{align*}

The extension to multiple Rx antennas is straightforward. The combined auxiliary variables become sums
\begin{align*}
	\breve{S}_{1j}	= 
  	   &\sum_{j=1}^{N_\text{Rx}}{R_j(m)H_{1j}^*} + \sum_{j=1}^{N_\text{Rx}}{R_j^*(m+1)H_{2j}}
	   \\
	   &\quad+ \sum_{j=1}^{N_\text{Rx}}{R_j^*(m+2)H_{3j}},
	\\
	\breve{S}_{2j} =
	   &\sum_{j=1}^{N_\text{Rx}}{R_j(m)H_{2j}^*} - \sum_{j=1}^{N_\text{Rx}}{R_j^*(m+1)H_{1j}}
	   \\
		&\quad + \sum_{j=1}^{N_\text{Rx}}{R_j^*(m+3)H_{3j}}, 
	\\
	\breve{S}_{3j}	=
	   &\sum_{j=1}^{N_\text{Rx}}{R_j(m)H_{3j}^*} - \sum_{j=1}^{N_\text{Rx}}{R_j^*(m+2)H_{1j}}
	   \\
		&\quad - \sum_{j=1}^{N_\text{Rx}}{R_j^*(m+3)H_{2j}}.
\end{align*}
which, in accordance with the one Rx antenna derivation, expand to
\begin{align*}
	\breve{S}_1
	&= \sum_{j=1}^{N_\text{Rx}}\sum_{i=1}^3 H_{ij}^*H_{ij}S_1
	+ \sum_{j=1}^{N_\text{Rx}}H_{1j}^*W_j(m)  		\\
	&\quad+ \sum_{j=1}^{N_\text{Rx}}H_{2j}W_j^*(m+1)
	 + \sum_{j=1}^{N_\text{Rx}}H_{3j}W_j^*(m+2),
\end{align*}
\begin{align*}
	\breve{S}_2
	&= \sum_{j=1}^{N_\text{Rx}}\sum_{i=1}^3 H_{ij}^*H_{ij}S_2
	- \sum_{j=1}^{N_\text{Rx}}H_{1j}W_j^*(m+1)  		\\
	&\quad+ \sum_{j=1}^{N_\text{Rx}}H_{2j}^*W_j(m)
	 + \sum_{j=1}^{N_\text{Rx}}H_{3j}W_j^*(m+3),
\end{align*}
\begin{align*}
	\breve{S}_3
	&= \sum_{j=1}^{N_\text{Rx}}\sum_{i=1}^3 H_{ij}^*H_{ij}S_3
	- \sum_{j=1}^{N_\text{Rx}}H_{1j}W_j^*(m+2)  		\\
	&\quad- \sum_{j=1}^{N_\text{Rx}}H_{2j}^*W_j(m+3)
	 + \sum_{j=1}^{N_\text{Rx}}H_{3j}W_j^*(m),
\end{align*}

Therefore, the minimum variance unbiased estimators of the transmitted constellation points for any number of Rx antennas become
\begin{equation*}
	\hat{S}_{1j} = 
	\frac{\breve{S}_1}{\displaystyle{\sum_{i=1}^3 H_{ij}^*H_{ij}}},
	\;
	\hat{S}_{2j} =
	\frac{\breve{S}_2}{\displaystyle{\sum_{i=1}^3 H_{ij}^*H_{ij}}},
	\;
	\hat{S}_{3j} =
	\frac{\breve{S}_3}{\displaystyle{\sum_{i=1}^3 H_{ij}^*H_{ij}}},
\end{equation*}
from which substitutions of $\breve{S}_1$, $\breve{S}_2$ and $\breve{S}_3$ give the final signal combiner (\ref{eq:EstTx3Orthogonal}) presented in Section \ref{sec:IVA}.

%
%
\section{SNR of the Combined Signal in the Orthogonal Code with three Tx Antennas}
\label{sec:AppendixB}
For the purpose of abbreviated notation, we assume that expectation of the constellation point energy is unity, i.e. $E\left( S^*S \right)=1$. This is scaling that does not affect the validity of the derivation in the general case. The corresponding expectation of the noise energy in the symbol estimate $\hat{S}_{1j}$ is given in eq. (\ref{eq:S1NoiseEnergy})
\begin{figure*} 
\begin{align}
	E\left[W\left(\hat{S}_{1j}\right)\right] 
	&= E\left[\left| \frac{W_j(m)H_{1j}^* + W_j^*(m+1)H_{2j} + W_j^*(m+2)H_{3j}}
	                      {\sum_{i=1}^3 H_{ij}^*H_{ij}}
	  \right|^2 \right] \nonumber \\
	&= E\left[\frac{H_{1j}^*H_{1j}W_j(m)^*W_j(m) +
	                H_{2j}^*H_{2j}W_j(m+1)^*W_j(m+1) +
	                H_{3j}^*H_{3j}W_j(m+2)^*W_j(m+2)}
                  {\left(\sum_{i=1}^3 H_{ij}^*H_{ij}\right)^2}
	  \right]
	\label{eq:S1NoiseEnergy}
\end{align}
\hrulefill
\end{figure*}
, which with the noise variance $\sigma^2$ further develops to
\begin{equation*}
	E\left[W\left(\hat{S}_{1j}\right)\right]
	= \frac{\left[\displaystyle{\sum_{i=1}^3 H_{ij}^*H_{ij}}\right] \sigma^2}
          {\left[\displaystyle{\sum_{i=1}^3 H_{ij}^*H_{ij}}\right]^2}
	= \frac{\sigma^2}
          {\displaystyle{\sum_{i=1}^3 H_{ij}^*H_{ij}}}.
\end{equation*}

Therefore, the signal-to-noise power ratio at the \emph{j}th Rx antenna becomes
\begin{equation*}
	\text{SNR}\left(\hat{S}_{1j}\right) = \frac{\displaystyle{\sum_{i=1}^3 H_{ij}^*H_{ij}}}
	                              {\sigma^2}.
\end{equation*}

Combining all Rx antennas does not change the expectation of the signal energy, which remains at unity. Provided the additive noise samples are independent and identically distributed (iid) among the Rx antennas, the expression for the corresponding combined noise energy becomes that of eq. (\ref{eq:CombinedNoiseEnergy}).
\begin{figure*} 
\begin{align}
	E\left[W\left(\hat{S}_1\right)\right] 
	&= E\left[\left| \frac{\displaystyle{\sum_{j=1}^{N_\text{Rx}} H_{1j}^*W_j(m)
	                                   + \sum_{j=1}^{N_\text{Rx}} H_{2j}W_j(m+1)^*
	                                   + \sum_{j=1}^{N_\text{Rx}} H_{3j}W_j(m+2)^*}}
	                {\displaystyle{\sum_{j=1}^{N_\text{Rx}}\sum_{i=1}^3 H_{ij}^*H_{ij}}}
	  \right|^2 \right] \nonumber \\
	&\quad = \frac{\displaystyle{\left[\sum_{j=1}^{N_\text{Rx}}\sum_{i=1}^3 H_{ij}^*H_{ij}\right]}}
           {\displaystyle{\left[\sum_{j=1}^{N_\text{Rx}}\sum_{i=1}^3 H_{ij}^*H_{ij}\right]}^2}
	 = \frac{\sigma^2}
           {\displaystyle{\sum_{j=1}^{N_\text{Rx}}\sum_{i=1}^3 H_{ij}^*H_{ij}}}
	\label{eq:CombinedNoiseEnergy}
\end{align}
\hrulefill
\end{figure*}

Hence, the signal-to-noise power ratio of the combined estimate is
\begin{equation*}
	\text{SNR}\left(\hat{S}_1\right) =
	\frac{\displaystyle{\sum_{j=1}^{N_\text{Rx}}\sum_{i=1}^3 H_{ij}^*H_{ij}}}
        {\sigma^2}.
\end{equation*}
which indeed is the sum of the constituent SNRs. The proof
showing that the combined SNRs of the estimates $\hat{S}_2$ and $\hat{S}_3$ are also the sums of their constituents is similar to the proof above.

%


\ifCLASSOPTIONcaptionsoff
  \newpage
\fi

%

\begin{IEEEbiography}{Risto Nordman}
(M'98) received the degree of B.Sc. in construction
engineering from Technical College of Kuopio, Finland, in 1978. After a
multifaceted career in the building branch he completely turned over and started
to study information technology, where he received the degrees of M.Sc. and
Lic.Tech. from the Department of Electrical Engineering, University of Oulu,
Finland, in 1997 and 1999, respectively.

In 1996 he joined VTT, where he currently holds a Research Scientist post
with Communications Platforms Research Area. He has a special expertise in
channel coding, and he has done a great deal of research work and project
managing in mobile telecommunication projects concerning e.g. UMTS and
WIMAX standards. Furthermore, he has participated in STINGRAY and
PULSERS EU projects. His current field of interest encompasses a larger view
on the PHY layer, from MIMO codes to channel estimation and further to FEC
coding aiming at improving the performance by joint design of the PHY layer
modules.
\end{IEEEbiography}

\vfill

\end{document}